\documentclass[intlimits,twoside,a4paper]{article}

\usepackage{amsmath,amssymb}
\usepackage{graphicx}
\usepackage{wrapfig}
\usepackage{color}

\usepackage[T2A]{fontenc}
\usepackage[cp1251]{inputenc}
\usepackage{cmpj2}

\issue{2013}{16}{3}{33703}
\doinumber{10.5488/CMP.16.33703}

\title[Density functional theory study of the $\alpha \to \omega$ martensitic transformation in titanium]{Density functional theory study of the $\alpha \to \omega$ martensitic transformation in titanium induced by hydrostatic pressure}%
\author[M.~Jafari \textsl{et al.}]{M.~Jafari\thanks{E-mail: Jafari@kntu.ac.ir} \ , M.~Nobakhti, H.~Jamnezhad, K.~Bayati}

\address{Department of Physics, K.N. Toosi University of Technology, Tehran, Iran}

\date{Received June 10, 2012,   in final form February 7, 2013}
\authorcopyright{M.~Jafari, M.~Nobakhti, H.~Jamnezhad, K.~Bayati, 2013}

\begin{document}

\maketitle

\begin{abstract}
The martensitic $\alpha \rightarrow \omega $ transition was investigated in Ti under hydrostatic pressure. The calculations were carried out using the density functional theory (DFT) framework in combination with the Birch-Murnaghan equation of state. The calculated ground-state properties of $\alpha $ and $\omega $ phases of Ti, their bulk moduli and pressure derivatives are in agreement with the previous experimental data. The lattice constants of $\alpha $ and $\omega$-phase at 0~K were modeled as a function of pressure from 0 to 74~GPa and 0 to 119~GPa, respectively. It is shown that the lattice constants vary in a nonlinear manner upon compression. The calculated lattice parameters were used to describe the $\alpha \rightarrow \omega $ transition and show that the phase transition can be obtained at 0~GPa and 0~K.

\pacs  71.15.Mb, 71.20.Be, 05.70.Ce, 64.70.K-
\end{abstract}

\section{Introduction}

Most of the elements of the periodic table display a transformation of their crystal structures under the application of temperature, pressure or both~\cite{1}. Group-IV transition metals, such as titanium, zirconium and hafnium, have attracted considerable interest due to their appealing electrical and mechanical properties. Their smaller density and higher strength make them the preferred metal of choice in many high-technology industries.

These electrical and mechanical properties may be changed by controlling the crystal structure. Such phase changes occur in pure titanium in response to the changes in pressure and temperature. The purpose of this work is to study its $\alpha \to \omega $ martensitic transition under hydrostatic pressure.  It should be noted that it is a scalar hydrostatic pressure rather than uniaxial mechanical stress.\textbf{  }

Martensitic transformation is a class of diffusionless, displacive and athermal transformations which is of great technological interest. The first-order martensitic transformations were considered in the present study where a unit cell of the parent phase was continuously distorted into a unit cell of the daughter phase. Certain vectors and planes in the parent and daughter phases were parallel and the transformation happened extremely quickly, giving the atoms insufficient time to travel very far.

At ambient pressure, the $\alpha $-phase (hexagonal close-packed structure or hcp) with space group P6${}_{3}$/mmc has two atoms per unit cell at the Wyckoff positions of (1/3, 2/3, 3/4) and (1/3, 2/3, 1/4)~\cite{2,3}. At high pressure, the $\omega $-phase (hexagonal structure) with the space group P6/mmm, has three atoms per unit cell and is composed of two planes, a hexagonal plane at $z = 0$ (A-plane) and a honeycomb plane at $z = 1/2c$ (B-plane). This is the AlB$_{2}$ type structure. Atoms are at (0, 0, 0) position in the A-plane and at (1/3, 2/3, 1/2), and (2/3, 1/3, 1/2) positions in the B-plane.

The $\alpha $-phase is known as the stable phase of Ti at room temperature and atmospheric pressure. By increasing the pressure, the $\alpha \to \omega $ transition was observed. At high temperatures and zero pressure, the $\omega $ phase transforms to    $\beta $-phase (body-centered cubic structure or bcc) below the melting point (1933~K)~\cite{4}. Except in Nguyan-Manh et al.~\cite{3}, theoretical calculations have shown that the $\omega $ phase is more stable than the $\alpha $ phase at 0~K~\cite{5,6,7,8,9}. Xia et al. showed that the $\omega $ phase remains stable up to 116~GPa~\cite{10}. Akahama et al.~\cite{11} reported a new $\eta $ phase with Fmmm space group in the range of $51\div80$~GPa with the $b/a$ ratio ranging from 0.704 to 1.007 and the $c/a$ ratio ranging from 0.704 to 0.789. Vohra et al.~\cite{12} investigated Ti-structures up to 146~GPa and obtained $\omega \to \gamma $ transformation at 116 $\pm$ 4~GPa. The $\gamma $-phase belongs to the orthorhombic lattices within the Cmcm space group. Akahama et al.~\cite{11} also obtained the $\omega \to \gamma $ phase transition, but they also reported for this transformation to occur at $124\div130$~GPa. Furthermore, they reported the transition from the $\gamma $-phase to a new $\delta$-phase at $140\div145$~GPa~\cite{11}.

In the experimental study with energy-dispersive X-ray diffraction (with a diamond anvil cell), the stability of different phases of Ti~\cite{10,12,13,14}, transition between $\alpha $ and $\omega $ phases~\cite{10,12,13,14,15}, the effects of impurity~\cite{12}, uniaxial stress and heating~\cite{14} on pressure-induced transition have been investigated. Moreover, the observed results of the X-ray diffraction show that the stability range of $\alpha $ phase varies between  the room temperature to around 923~K~\cite{16}.

All theoretical works on transformation have been based on finding the minimum energy or Gibbs free energy. To obtain the Gibbs free energy for each structure, researchers have used different methods such as the linear Muffin-Tin orbital and the local-density approximation (LDA) methods within the framework of the DFT~\cite{5}, the full-potential linearized-augmented-plane wave (FP-LAPW) method and LDA~\cite{3,13}, the FP-LAPW and generalized-gradient approximation (GGA)~\cite{7}, first-principles energy calculations~\cite{17}, the tight-binding total energy (TBE) method with first-principles calculations~\cite{18}, the epitaxial bain path (EBP) combined with FP-LAPW and GGA~\cite{9} and the modified embedded-atom  potential method and DFT~\cite{19}.

\section{Methodology }

This study is solely concerned with the $\alpha $ and $\omega $ structures of Ti. The Gibbs free energy per atom obtained as a function of the volume in the two phases was fitted to the Birch-Murnaghan's equation of state (EOS). At zero temperature, the calculations of total energy were carried out using the FP-LAPW method supplemented by local orbitals (FP-LAPW+lo) method based on the DFT~\cite{20}. The exchange and correlation functionals were given by the GGA of Perdew-Wang~\cite{21,22}. The calculations were carried out using the WIEN2K package~\cite{20}. It is well known that the application of the FLAPW method to metallic-like compounds always underestimates the inter-band energy distances~\cite{23} and very often a scissors operator is used~to shift the conduction bands in order to achieve an agreement with the experimental value.

Since the bulk Ti exhibits paramagnetic properties, the spin effects were neglected. The value of   $R_\textrm{MT}K_\textrm{max} = 8.5$ obtained after optimization was selected for both $\alpha $ and $\omega $, where RMT is the average radius of the muffin-tin spheres and $K_\textrm{max}$ is the maximum value of the wave vector~$K$.

The optimized $R_\textrm{MT}$ for the highest pressure was calculated and it was used for the remaining pressures in each phase. Although this was not necessary for low pressures (because the muffin-tin spheres did not intersect at these pressures), it facilitated the subsequent comparison of the results from various pressures. The radii of the spheres used in this study were 2.208~{\AA} for the $\alpha $ phase and 2.300~{\AA} and 2.060~{\AA} for the $\omega $ phase.

A desirable charge convergence was determined as $1 \times 10^{-4} \, e$ and the largest vector in the charge-density Fourier expansion was fixed at $G_\textrm{max}=12$~bohr$^{-1}$.

Following the $k$-point optimization at $P = 0$, 731 points were found for the hcp and 2460 for the simple hexagonal structures, which corresponded to 64 and 171 $k$-point mesh in the irreducible wedge of the Brillouin zone, respectively.

In the first step, $k$-mesh was optimized at the equilibrium volume. For each volume smaller than $V_{0}$, the same $k$-point as that generated for the zero-pressure case was used. The hexagonal structure could be described by two parameters ${a}$ and ${c}$ (which included the axial ratio ${c/a}$). The following procedure outlines the current method used to optimize the ${c/a}$ ratio at each phase and volume. In the first step, by using the WIEN2K code, the total energy was calculated for several values of ${c/a}$ and keeping the volume ($V$) fixed. Then, the obtained results were fitted to a polynomial form in order to find the equilibrium value of the ${c/a}$ ratio. The minimum energy related to the minimum value of ${c/a}$ in the polynomial form was compared with the energy given by WIEN2K. If the energy related to the minimum ${c/a}$ in the polynomial form was the minimum energy provided by WIEN2K, it was accepted; otherwise, this procedure was repeated with other values of ${c/a}$ until the minimum energy was found. This procedure was repeated for various volumes. To investigate the thermodynamic properties of the considered structures, the following form representing the third-order Birch-Murnaghan EOS was used:
\begin{equation}
E(V)=E_0+\frac{9B_0V_0}{16}\left\{\left[\left(\frac{V_0}{V}\right)^{\frac{2}{3}}-1\right]^3
+ \left[\left(\frac{V_0}{V}\right)^{\frac{2}{3}}-1\right]^2
\left[6-4\left(\frac{V_0}{V}\right)^{\frac{2}{3}}\right]\right\},
\end{equation}
where $E_0$ and $V_0$ are the energy and volume at equilibrium, respectively, $B_0$ is the zero-pressure bulk modulus and $B'_{0}$ is the pressure derivative of $B_0$. For all cases, $B_{0}$ and $B'_{0}$ can be determined by fitting the $E (V)$ points from DFT to the Birch-Murnaghan's EOS. Moreover, it was found that for each volume, the pressure depends on the third-order Birch-Murnaghan's EOS as follows:
\begin{equation}
P(V)=\frac{3B_0}{2}\left[\left(\frac{V_0}{V}\right)^{\frac{7}{3}}- \left(\frac{V_0}{V}\right)^{\frac{5}{3}}\right]
\left\{1+\frac{3}{4}(B_0-4)\left[\left(\frac{V_0}{V}\right)^{\frac{2}{3}}-1\right]\right\}.
\end{equation}
The theoretical pressure dependences of the lattice constants and the axial ratios ${c/a}$ can be calculated.

\begin{figure}[!b]
\centerline{
\includegraphics[width=0.48\textwidth]{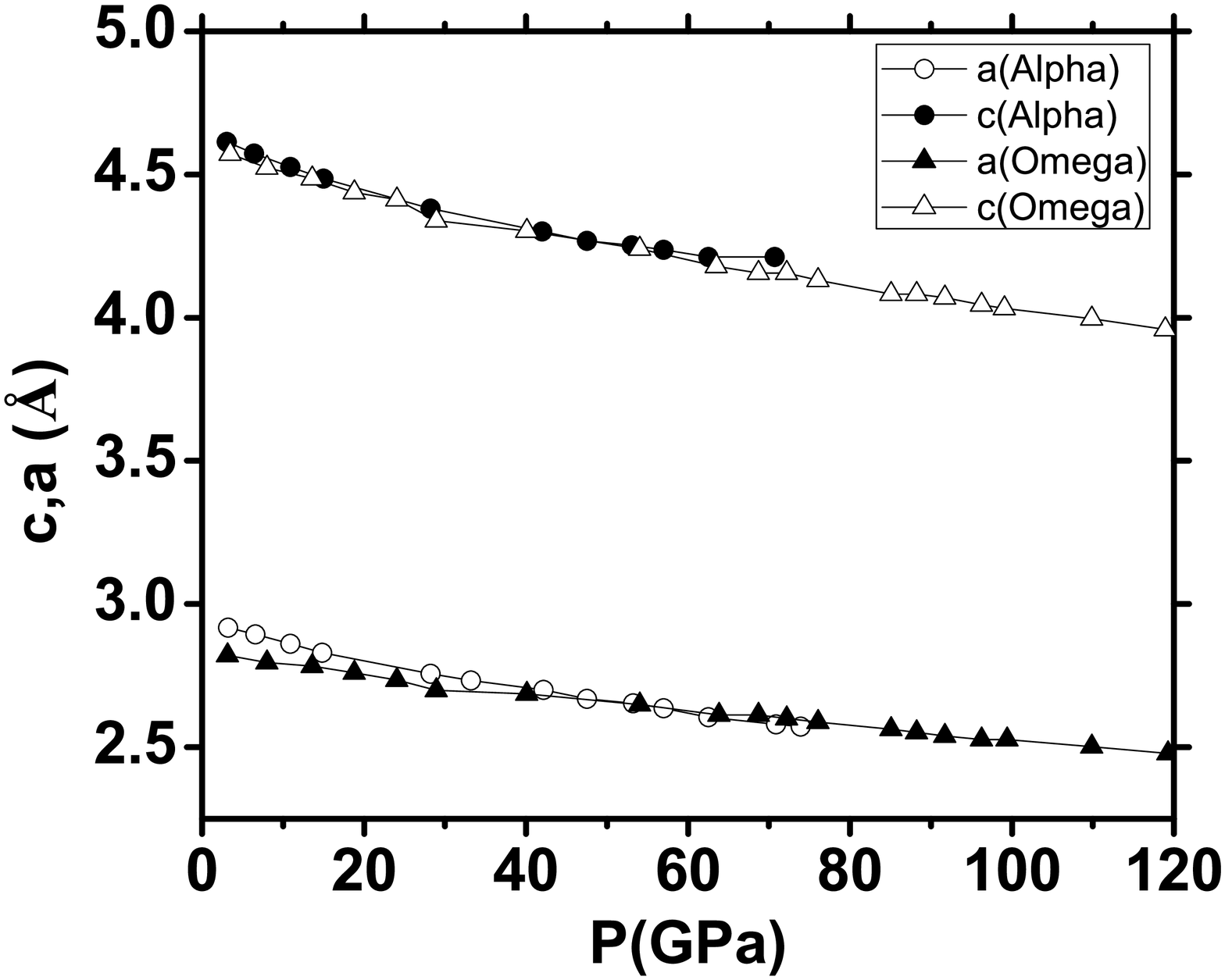}
\hspace{1mm}
\includegraphics[width=0.5\textwidth]{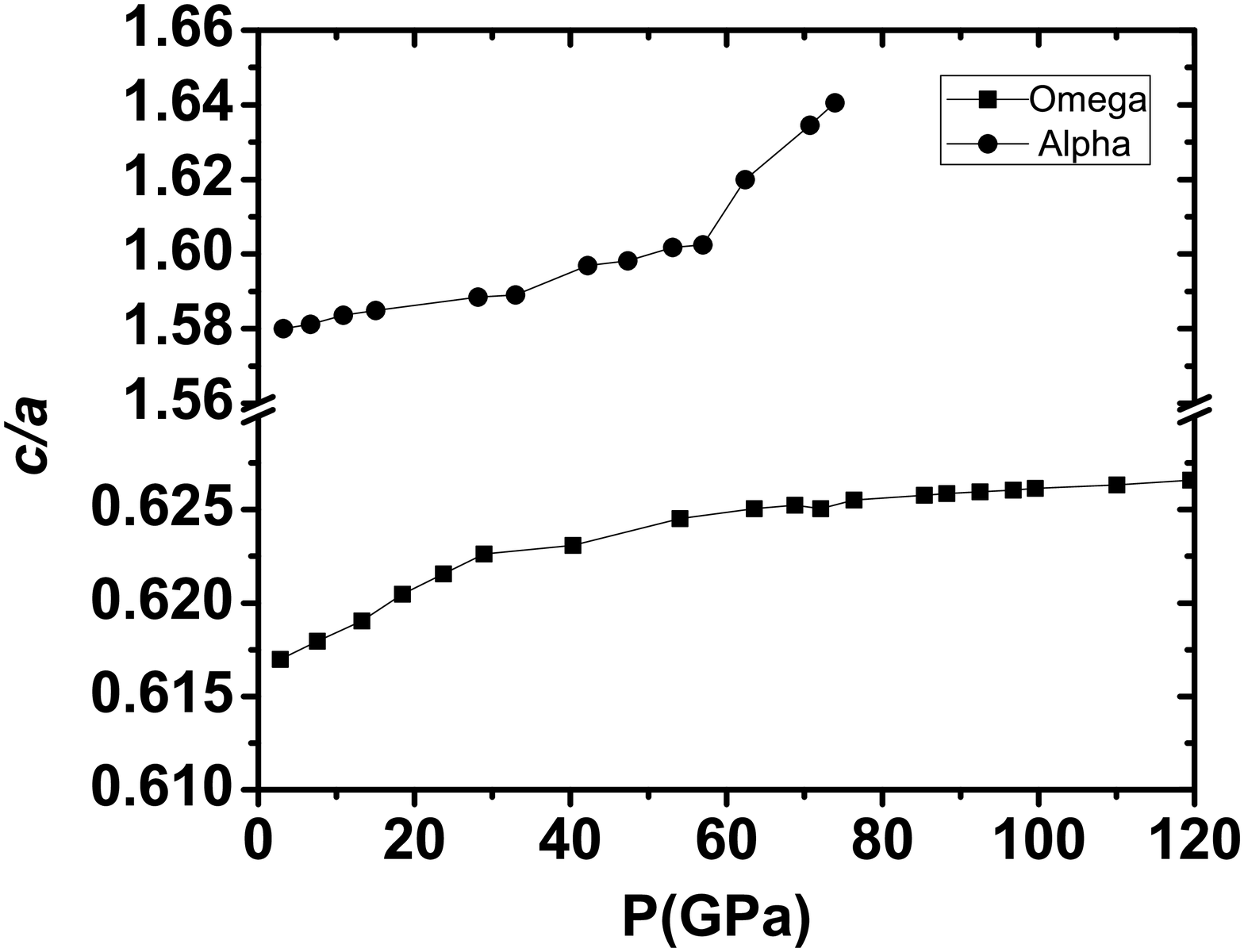}
}
\parbox[t]{0.5\textwidth}{
\caption{Theoretical unit cell parameters vs. pressure for the $\alpha$ and $\omega $-Ti (lines are for the eye's guide).\label{fig1}}
}
\parbox[t]{0.5\textwidth}{
\caption{Theoretical ${c/a}$ ratio vs. pressure for the $\alpha$ and $\omega $-Ti (line is for the eye's guide).\label{fig2}}
}
\end{figure}

The lattice parameters of two phases are shown in figure~\ref{fig1}. For both phases a nonlinearity was observed with an increasing pressure. The $c/a$ ratios in both phases increased upon compression, but in a different manner (see figure~\ref{fig2}). In the $\alpha$ phase, there was a positive concavity and it was negative in the $\omega$ phase.

The Gibbs free energy was derived using the equation, $G = H-TS$, where $H$ stands for the enthalpy of a system. At $T = 0$~K, the enthalpy is equal to the Gibbs free energy and thus $G = H= E + PV$. The transition pressure was obtained by comparing the enthalpies of the two phases.

\section{Results and discussion}

\begin{figure}[!t]
\centerline{
\includegraphics[width=0.5\textwidth]{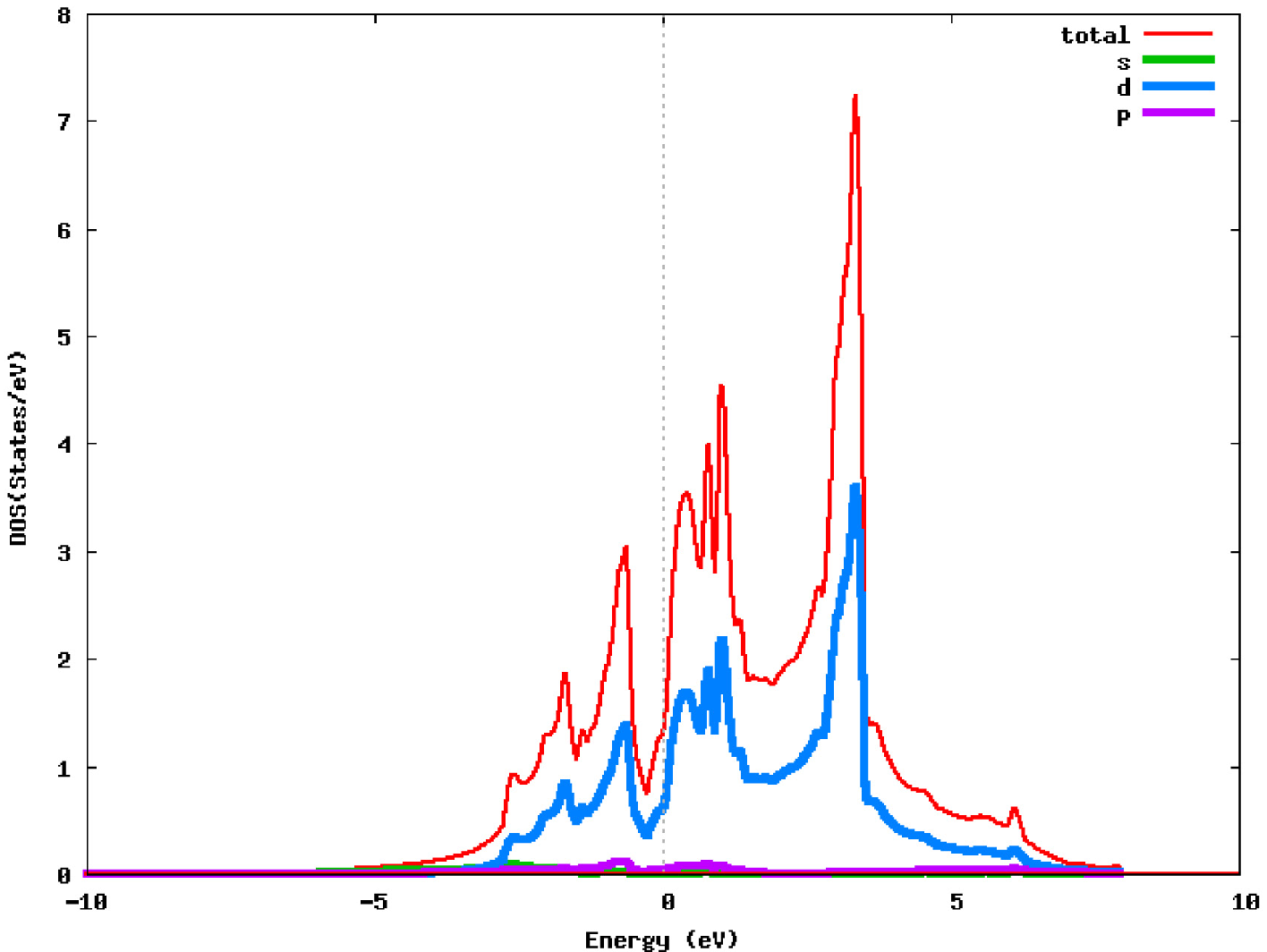}
\includegraphics[width=0.5\textwidth]{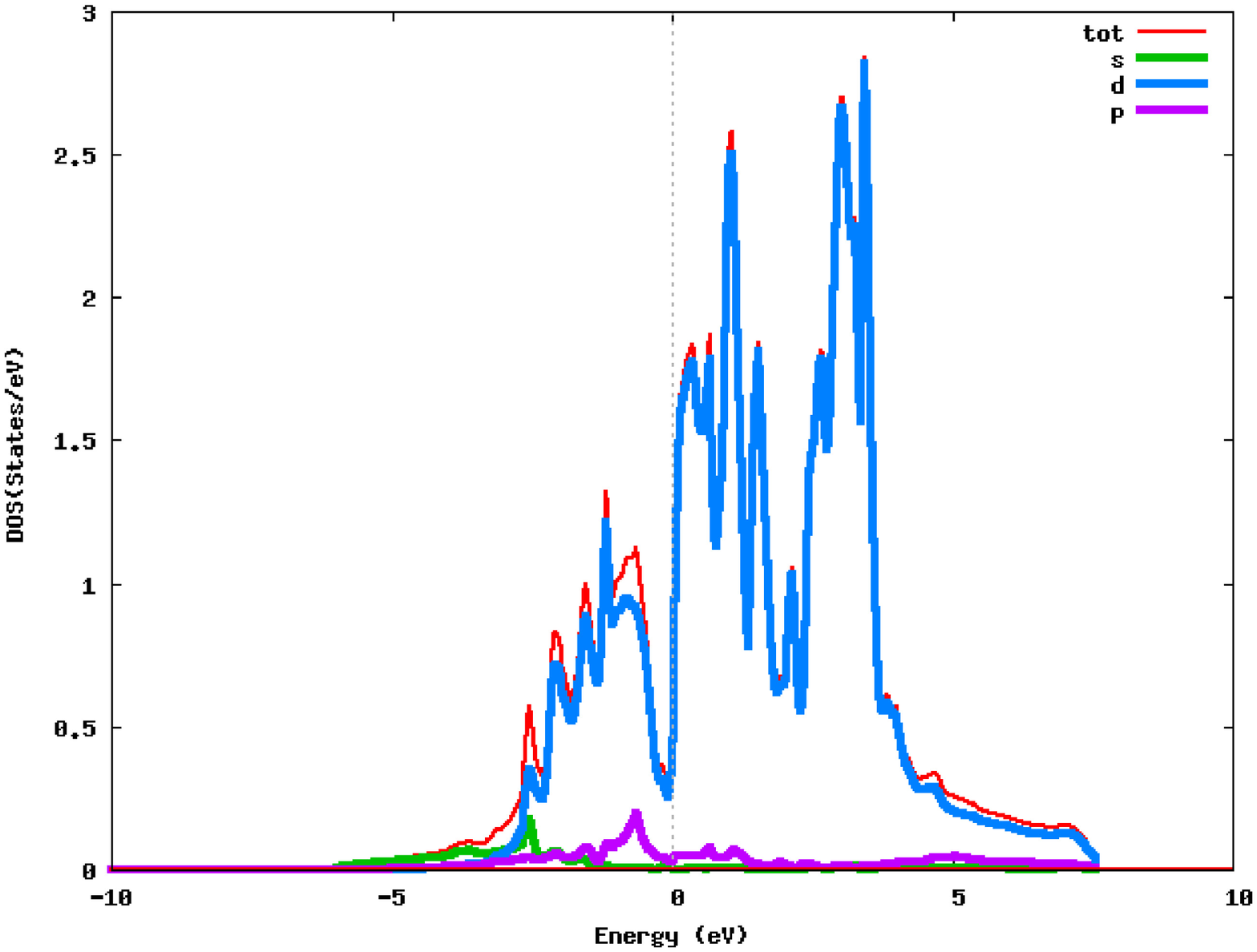}
}
\parbox[t]{0.5\textwidth}{
\caption{(Color online) Total and partial DOS of the $\alpha$-Ti.\label{fig3}}
}
\parbox[t]{0.5\textwidth}{
\caption{(Color online) Total and partial DOS of the $\omega$-Ti.\label{fig4}}
}
\end{figure}

The electronic densities of states (DOS) for the $\alpha $ and $\omega $ phases are shown in figures~\ref{fig3} and~\ref{fig4}. As can be seen, the Fermi energy level ($E_\textrm{f}$) is at the minimum of the DOS curves in both phases. Furthermore, the minimum around $E_\textrm{f}$ in the $\omega $ phase is narrower than that in the $\alpha$ phase. These features are in good agreement with the previous works~\cite{24}. The total energy of $-1707.6354$~Ryd/atom was calculated for $\alpha $ and $-1707.63600$~Ryd/atom for $\omega $ phases at $T=0$~K and $P=0$~GPa. In our previous work~\cite{25}, the electrical properties were determined from the location of the Fermi level and the DOS per atom at $N$ ($E_\textrm{f}$). Furthermore, in both phases, the $d$-states dominate at $N$ ($E_\textrm{f}$)~\cite{26}.

\begin{table}[htb]
\caption{The optimized lattice constants, $B_0$, $B'_{0}$ at $T = 0$~K and $P = 0$~GPa for the $\alpha $ and $\omega $-Ti.}
\label{tab1}
\vspace{1ex}
\begin{center}
\begin{tabular}{|c|l|l|l|l|}
\hline
&  \multicolumn{2}{|c|}{$\alpha $  phase} &  \multicolumn{2}{|c|}{$\omega $  phase} \\
\hline
& This work & Experiment &   This work & Experiment \\
\hline\hline
$B_0$ & $101 \pm 6$~GPa & $117\pm 9$~GPa~\cite{28},   &
$137 \pm  3$~GPa & $138 \pm 10$~GPa~\cite{28} \\
&  & $117.4$~GPa~\cite{29} & &  \\
\hline
$B'_{0}$ & 3.59$\pm 0.02$ & 3.9${  \pm }{  0.4}$~\cite{28}, &   2.94$\pm 0.04$ & 3.8${  \pm }{  0.5}$~\cite{28} \\
&  & 3.444~\cite{29} & &  \\
\hline
$a$ & 2.948017~{\AA} & 2.956$\pm$0.002~{\AA}~\cite{29} &   4.591137~{\AA} & 4.598 $\pm$  0.009~\cite{12} \\
\hline
$c$ & 4.650202~{\AA} & 4.734$\pm$0.025~{\AA}~\cite{29} &  2.834109~{\AA} & --- \\
\hline
$c/a$ & 1.57740 & 1.601~\cite{29} &   0.61730 & 0.614~\cite{12} \\
\hline
\end{tabular}
\end{center}
\end{table}

\begin{wrapfigure}{i}{0.5\textwidth}
\centerline{
\includegraphics[width=0.49\textwidth]{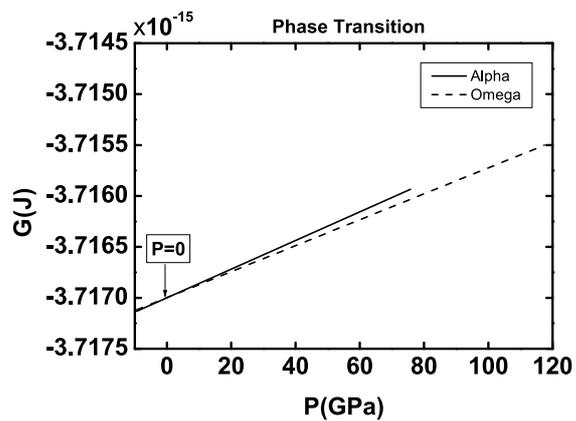}
}
\caption{ Gibbs free energies as a function of pressure for $\alpha$ and $\omega $ phases of Ti.\label{fig5}}
\end{wrapfigure}
%

This indicates that the chemical bonding of the hexagonal structure is quite different from that expected for transition metals. The hexagonal structure shows a large degree of covalency in the $B$ site with $sd^{2}$ hybridization. These properties cause more hardiness and low conductivity due to some electrons localized in the hybrid orbital (Doherty and Gibbons model). It should be noted that the anharmonic electron-phonon interaction also plays an important role in the pressure induced phase transition~\cite{27}.

The optimized lattice constants and the values of $B_0$ and $B'_{0}$ are given in
table~\ref{tab1}. The optimized theoretical lattice parameters at zero pressure in the $\alpha $-phase were within 95\% of the experimental data. The pressure range for the $\alpha $-phase and the $\omega$-phase were $0\div74$~GPa and $0\div119$~GPa, respectively. The results are presented in figures~\ref{fig1} and~\ref{fig2}.

Least-squares curve fitting was used to determine a model for these data. It was found that a linear model was not capable of accurately capturing the parameter variations, and instead a nonlinear model was evaluated. Nonlinear behavior was visible in all the parameters (${a}$, ${c}$, and ${c/a}$).

Moreover, the thermodynamic parameters ($E,V$) calculated using the DFT theory were fitted to the third-order Birch-Murnaghan's  EOS and then the Gibbs free energy was obtained with these three parameters ($E,V$ and $P$).The transition pressures were found by comparing the enthalpy values of the two phases. Furthermore, our results show that these two Gibbs free energy curves (figure~\ref{fig5}) intersect at 0~GPa and 0~K.

The extrapolation of experimental data~\cite{30} predicted that a martensitic $\alpha \to \omega $ transition occurs near $-1$~GPa at 0~K. Our results are within the hysteresis interval ($-5$ to $+8$~GPa) and remain in excellent agreement with the other experimental and theoretical works~\cite{9,18,31}. Theoretical investigations~\cite{9} were carried out using first-principles total-energy calculations and the EBP for finding the equilibrium structures directly as a function of the pressure from the minima of the particular Gibbs free energies. These studies reported that the $\alpha$-phase was stable up to 50~GPa and it transformed to the $\omega $ structure at 50~GPa. In another theoretical work~\cite{18}, the $\alpha \to \omega $ transition in Ti was studied using the TBE method and it was reported that the $\alpha \to \omega $  transition in Ti occurred at 6~GPa. First-principles total-energy calculations with GGA approximation were also used to investigate the $\alpha \to \omega $ transition in Ti~\cite{31}. They compared the enthalpy values of different structures and obtained $-3$~GPa for the $\alpha $ to $\omega $ transition. These results are summarized in table~\ref{tab2}.

\begin{table}[htb]
\caption{Calculated pressure-induced $\alpha \to \omega $ transition.}
\label{tab2}
\vspace{1ex}
\begin{center}
\begin{tabular}{|c|r|}
\hline
Theoretical work & Pressure  \\
\hline\hline
\cite{9}  &  $50$~GPa\\
\hline
\cite{18} &  $6$~GPa \\
\hline
\cite{31} &  $-3$~GPa \\
\hline
This work &  $0$~GPa \\
\hline
\end{tabular}
\end{center}
\end{table}

Based on the comparison between this paper's calculated pressure-induced transition and the extrapolated data from experiments, it can be suggested that this method can also be used for explaining the structural sequence in Ti under pressure for other phases.

\section{Conclusion}

This study investigated the martensitic $\alpha $ to $\omega $ transition in Ti under hydrostatic pressure. The calculations were carried out using FP-LAPW+lo method. Accordingly, the lattice constants, $c/a$ ratios, electronic structures, the bulk moduli and pressure derivatives of the two Ti phases were calculated. Thereby, it was shown that most of the structural parameters vary in a nonlinear manner. The calculated parameters were used for describing the $\alpha \to \omega $ martensitic transition under hydrostatic pressure and showing that this transition was obtained at 0~GPa and at zero temperature. This value was extremely close to the extrapolated experimental data.

\ukrainianpart

\title{Дослідження методом функціоналу густини мартенситового переходу ${\mathbf \alpha}$ в ${\mathbf \omega}$ в титані під дією гідростатичного тиску}

\author{M.~Джафарі, M.~Нобахті, H.~Джамнезхад, K.~Баяті}

\address{Факультет фізики, Технологічний  університет ім. К.Н. Тусі,
Тегеран, Іран}

\makeukrtitle

\begin{abstract}
\tolerance=3000%
Досліджувався мартенситовий перехід  $\alpha \rightarrow \omega $ в
Ti при гідростатичному тиску. Обчислення здійснювалися із
використанням теорії функціоналу густини (DFT) в комбінації з
рівнянням стану Берча-Мурнагана. Обчислені властивості основного
стану $\alpha $ та $\omega $ фаз Ti, їх об'ємні модулі і тискові
похідні узгоджуються із попередніми експериментальними даними.
Постійні ґратки $\alpha $ і $\omega$-фази при 0~K моделювалися як
функція тиску, що змінювався від 0 до 74~GPa та від 0 до 119 GPa, відповідно. Показано,
що ґраткові сталі змінюються з концентрацією нелінійно. Обчислені
ґраткові параметри використовувалися для опису переходу $\alpha
\rightarrow \omega $ і було показано, що фазовий перехід може бути
отриманий при 0 GPa та 0 K.

\keywords титан, мартенситовий перехід, гідростатичний тиск, вільна
енергія Гіббса, DFT обчислення

\end{abstract}

\end{document}